\documentclass[journal]{IEEEtran}
\usepackage{graphicx, dblfloatfix}
\usepackage{amsmath}
\usepackage{amssymb}
\usepackage{graphics}
\usepackage{cite}
\usepackage{amsfonts}
\usepackage{textcomp}
\usepackage{multicol}
\usepackage{multirow}
\usepackage{color}
\usepackage{fixltx2e}
\usepackage{epstopdf}
\usepackage{bm}
\usepackage{siunitx}
\usepackage{algorithm}
\usepackage{algorithmic}
\usepackage{enumitem}
\usepackage{paralist}
\usepackage[dvipsnames]{xcolor}

\begin{document}

\title{An Endless Optical Phase Delay for Phase Synchronization in High-Capacity DCIs}

\author{Rakesh Ashok, Sana Naaz, and Shalabh Gupta\vspace{-10pt}

 \thanks{The authors are with the Department of Electrical Engineering, Indian Institute of Technology Bombay, 
 Mumbai -- 400076, India (email: rakesh.ashok@iitb.ac.in; sananaazsana@ee.iitb.ac.in; and shalabh@ee.iitb.ac.in).
 R. Ashok and S. Naaz have equally contributed to this work.
}}

\markboth{}%
{Shell \MakeLowercase{\textit{et al.}}: Bare Demo of IEEEtran.cls for IEEE Journals}

\maketitle

\begin{abstract}
In this work, we propose and demonstrate a module to linearly add an arbitrary amount of continuous (reset-free) phase delay to an optical signal. The proposed endless optical phase delay (EOPD) uses an optical IQ modulator and control electronics (CE) to add the desired amount of phase delay that can continuously increase with time. In order to adjust for the bias voltages and control voltage amplitudes in the EOPD, some of which may be time varying, a multivariate gradient descent algorithm is used. The EOPD has been 
demonstrated experimentally, and its use in a high-capacity data center interconnect (DCI) application has been outlined in this letter. 
The EOPD may find its use in many other applications that require precise phase/frequency adjustments in real-time. 

\end{abstract}

\begin{IEEEkeywords}
Continuous phase modulation, coherent communication, electro-optic phase modulator, gradient descent.
\end{IEEEkeywords}

\IEEEpeerreviewmaketitle

\section{Introduction}

\IEEEPARstart{E}{lectro-optic} phase modulators (PMs) are essential components in optical communication links, optical wireless access networks,
and radars and beam-forming antenna phased arrays based transceivers \cite{Ho_PMOCS2005,Yu_OFC2006,Hietala_OTMA1991}. 
The PMs in these systems are used for (but not limited to) generating optical phase shift, phase synchronization, frequency comb generation, and
frequency stabilization of lasers \cite{Wu_OL2010,Ji_PTL2012,Ashok_OFC2019}. 
Depending on the application, PMs are evaluated based on the performance metrics that
include drive voltage, half-wave voltage,
electrical modulation bandwidth, optical bandwidth, dynamic
range, extinction ratio (modulation depth), and linearity \cite{Hunsperger_IOTT2009}.
Various PMs with high-performance parameters such as high bandwidth, high linearity, high modulation efficiency, 
and low-loss have been discussed in the literature. However, the maximum amount of phase change provided 
by these conventional PMs is limited due to constraints on the magnitude of the electrical drive signal that can be applied and material properties of the modulator as they typically follow the relation: $\theta_{out}=\pi V_{drive}/V_{\pi}$, wherein $\theta_{out}$, $V_{drive}$, and
$V_{\pi}$ represent the phase of the output signal, applied drive voltage, and half-wave voltage of the PM, respectively. 
{If
a tunable phase shift of upto $2\pi$ can be provided by the
PM, any phase shift can be achieved in theory. However, if the phase shift has to be changed continuously with time, 
these PMs cannot be used}. 

A few approaches have been proposed in the literature for generating an infinite amount of phase delay {over the time, which is called endless (or boundless) phase delay}. 
A demonstration shown in \cite{Madsen_JLT2006} is
implemented on an integrated platform comprising multiple phase shifters and directional couplers in a cascade, which requires control signals that are difficult to synthesize for high-speed error-free phase delay generation. 
Another architecture, described in \cite{Doerr_USPatent2014}, uses a series of Mach-Zehnder interferometer (MZI) switches, 
comprising phase shifters and a line phase shifter along with multiple analog control signals to generate the desired phase shift. Here, the main limitation is the requirement of a large number of components and electrical signals that have to be controlled precisely.
The phase delay generator demonstrated in \cite{Ozharar_PTL2005}, uses the principle of serrodyning with time shift and phase shift approaches. However, the system exhibits large relative phase fluctuations due to temperature variations and requires precise control of optical and electrical signals. Also, it requires switching of the optical signal, which makes it less suitable for practical applications.

In this work, we propose an optical IQ modulator based module, along with a methodology to generate control signals, to achieve an EOPD that can be used in practical applications. The proposed module overcomes most of the limitations associated with the existing endless phase delay {techniques}. To overcome the non-idealities of the IQ modulator, an optimization procedure has been proposed and detailed. Experimental characterization of the EOPD and its use for phase offset correction in a polarization multiplexed carrier based self-homodyne (PMC-SH) link have also been presented.

\section{EOPD Architecture}

\begin{figure}[t!]
    \centering
    \vspace{-.1cm}
    \includegraphics[width=.38\textwidth]{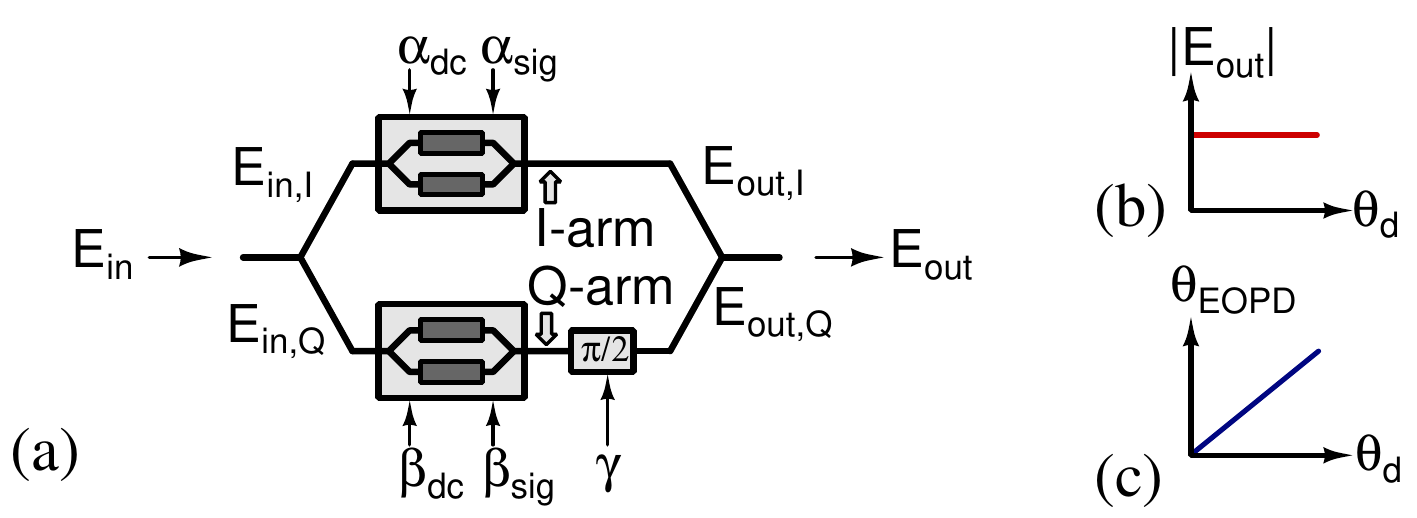} 
    \caption{(a) Architecture of an endless optical phase delay comprising an optical IQ modulator and control electronics; 
    Necessary conditions for endless optical phase delay operation: (b) Magnitude condition; and 
    (c) Phase condition.}
    \label{fig:EOPD_structure}   
  \end{figure}

\begin{figure}[t!]
\vspace{-.05cm}
    \centering
    \includegraphics[width=.33\textwidth]{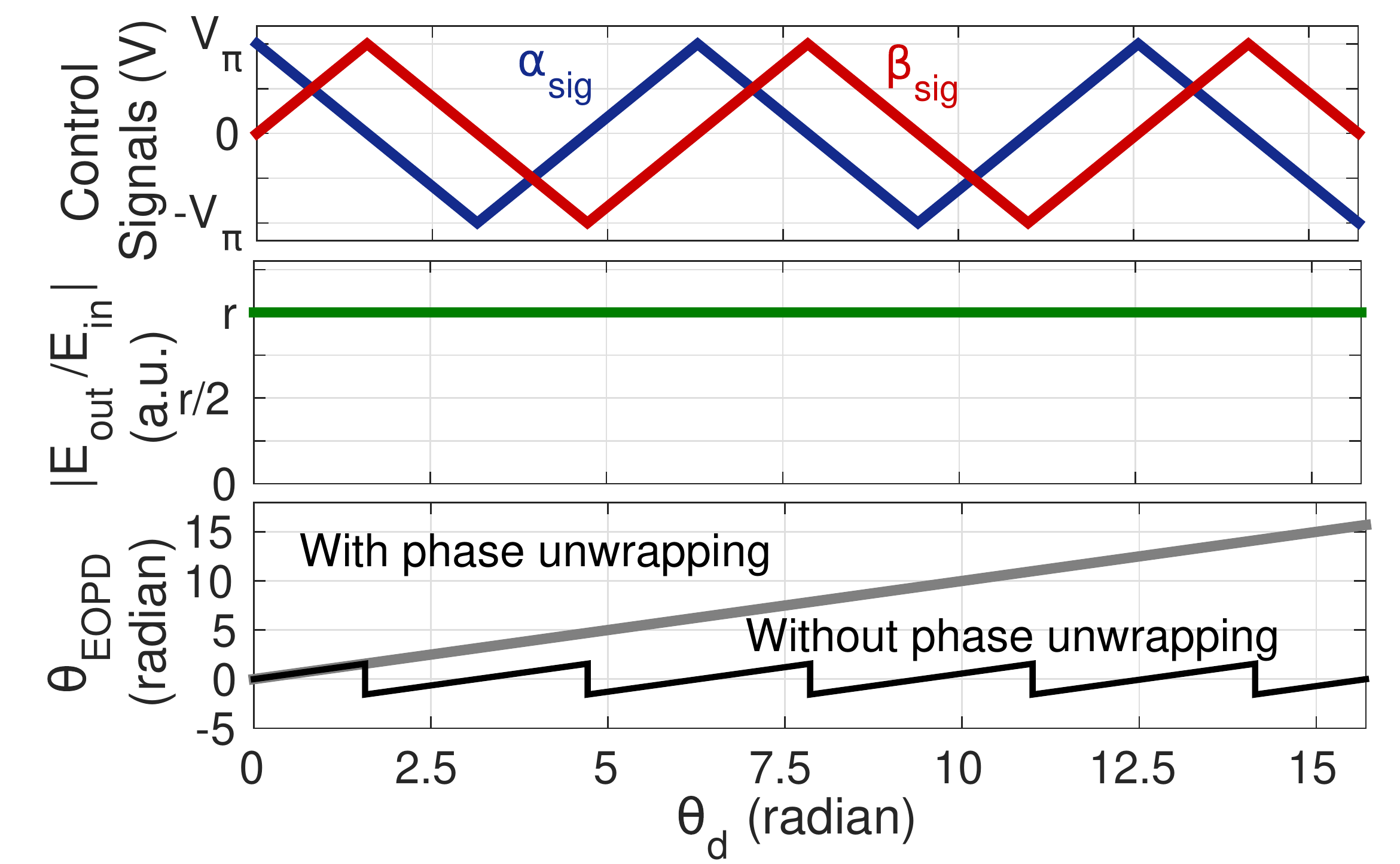} 
    \caption{{Simulation results of an endless optical phase delay. $\alpha_{sig}$, $\beta_{sig}$: Control signals; $E_{in}$, 
    $E_{out}$: Input and output electric fields; $\theta_d$ and $\theta_{EOPD}$: Desired phase shift and resultant phase shift.}}
    \label{fig:EOPD_sim_results}   
  \end{figure}

 {
\noindent The proposed EOPD, shown in Fig. \ref{fig:EOPD_structure}(a), comprises an IQ modulator, having two Mach-Zehnder modulators (MZMs) (one each in the I and the Q arms) and
a PM embedded in the Q-arm, and CE. The CE generate
the bias voltages ($\alpha_{dc}$, $\beta_{dc}$, and
$\gamma$) and control signals ($\alpha_{sig}$ and $\beta_{sig}$), and actively adjust them to compensate for the dynamic 
 variations. The EOPD optical input  $E_{in}=E_o\exp(j\omega_ot)$ has magnitude $E_o$ and 
frequency $\omega_o$, and generates output $E_{out}=E_o\exp[j(\omega_ot+\theta_{EOPD})]$, 
which is the phase delayed version of the input (with a phase shift $\theta_{EOPD}$). To achieve the desired phase shift $\theta_d$, that has to be changed with time without any discontinuity, $\alpha_{sig}$ and $\beta_{sig}$ are applied as continuous functions of time.}
{
For simplicity, we assume that each of the internal MZMs is biased at its null point, and the PM is biased to provide a phase shift of $\pi/2$. With this assumption, we can write the following expression 
(without explicitly showing the biases $\alpha_{dc}$, $\beta_{dc}$, and $\gamma$): 
\begin{align} \label{eqn:eout}
\begin{split}
\dfrac{E_{out}}{E_{in}} & =  \sin\left(\dfrac{\pi\alpha_{sig}}{2V_{\pi}}\right)  +j\sin\left(\dfrac{\pi\beta_{sig}}{2V_{\pi}}\right),  
\end{split}
\end{align}
where $V_{\pi}$ is the half-wave voltage of each of the MZMs. 
The phase added by the EOPD, i.e. $\theta_{EOPD}~ (=\angle{E_{out}}-\angle{E_{in}})$ can be obtained from (1) and written as: 
\begin{align}
\theta_{EOPD}=\arctan\left[\dfrac{\sin\left({\pi\beta_{sig}}/{2V_{\pi}}\right)}{\sin\left({\pi\alpha_{sig}}/{2V_{\pi}}\right)}\right] + n\pi, 
\label{eqn:eout_phase}
\end{align}
where, $n$ is the integer that unwraps the phase correctly \cite{oppenheim2016signals}. Phase unwrapping ensures that $\theta_{EOPD}$ is a continuous function of time when $\alpha_{sig}$ and $\beta_{sig}$ are also continuous functions of time. The operating constraints for the EOPD are: \begin{inparaenum}[(i)]
            \item the magnitude constraint, i.e. the magnitude of $|E_{out}/E_{in}|$ has to be constant and $\leq 1$ (say $r$), irrespective of the value of $\theta_{d}$, as shown in Fig. \ref{fig:EOPD_structure}(b) and represented by using:
            \begin{align}
 \sin^2\left(\dfrac{\pi\alpha_{sig}}{2V_{\pi}}\right)+\sin^2\left(\dfrac{\pi\beta_{sig}}{2V_{\pi}}\right)=r;
 \label{eqn:eout_mag}
\end{align}               
            and
            \item the phase constraint set by (\ref{eqn:eout_phase}).
        \end{inparaenum}
Considering that $\theta_{EOPD}$ is ideally equal to $\theta_d$, and using (\ref{eqn:eout_phase}) and (\ref{eqn:eout_mag}), we obtain
$\sin(\pi\alpha_{sig}/{2V_{\pi}})=r\cos\theta_d$ and $\sin(\pi\beta_{sig}/{2V_{\pi}})=r\sin\theta_d$. It is important to note that multiple values of $\alpha_{sig}$ and $\beta_{sig}$ can satisfy these constraints. However, we choose the values that keep $\alpha_{sig}$ and $\beta_{sig}$ real, bounded, and continuous versus time, so that $\theta_{EOPD}$ is also continuous with time. For $r=1$, 
 $\alpha_{sig}$ and $\beta_{sig}$ turn out to be triangular waveforms, that have peak-to-peak amplitudes of $2V_{\pi}$ and are delayed by one-fourth period relative 
 to each other, as shown in simulation results (Fig. \ref{fig:EOPD_sim_results}). 
{The magnitude transfer function in (\ref{eqn:eout_mag}) is constant ($=r$), while $\theta_{EOPD}$, monotonically increases 
with $\theta_d$ with phase unwrapping.}
{An EOPD adds a phase delay to $E_{in}$ with a slope of $2\pi f_{con}$ ($f_{con}$: frequency of 
control signals). Each cycle of control signals adds $2\pi$\,radians (Fig. \ref{fig:EOPD_sim_results}) 
of phase shift to $E_{in}$ and with $k$ cycles, $2k\pi$\,radians of 
phase shift can be achieved for any real value of $k$.}
In order to validate the magnitude 
constraint, $E_{out}$ signal is given to a photodetector (PD) resulting in photocurrent proportional to $|E_{out}|^2$ 
{(represented by $M_1$)},
which is required to be constant with varying $\theta_{d}$. For validating the phase condition, interferometer structure comprising 
a coupler and a PD that acts as a photo-mixer is used. The output of the coupler is $E_c=E_{in}+E_{out}$ and the 
corresponding photocurrent, is required to be proportional to $2E_o^2[1+\cos(\theta_{d})]$ {(denoted by $M_2$)} 
and has negligible distortion.}

\section{EOPD: Optimization and characterization}
\noindent In order to bias the two MZMs of the IQ modulator at their null points, 
suitable biases $\alpha_{dc}$ and $\beta_{dc}$ have to be added to $\alpha_{sig}$ and $\beta_{sig}$, respectively. 
Also, to provide a $\pi/2$ phase shift between optical signals in I and Q arms, a suitable bias $\gamma$ has to be provided 
(as shown in Fig. \ref{fig:EOPD_structure}(a)). 
To correct dynamic variations in the properties of the IQ modulator (that cause drift in $\alpha_{dc}, \beta_{dc}$, and $\gamma$), the use of a multivariate iterative gradient descent algorithm has 
been proposed and demonstrated. 
 We use the normalized magnitude condition $M_1=\Re|E_{out}|^2$ as the input, with $M_{1,A}$ and  $M_{1,P}$ as the 
actual and predicted magnitudes, respectively. The risk function 
$\mathbf{\textit{J}}$, which is the mean square difference between $M_{1,A}$ and  $M_{1,P}$ is minimized by optimizing the 
EOPD parameters (bias voltages: $\alpha_{dc}$, $\beta_{dc}$, and $\gamma$; and 
control voltage amplitudes: $\alpha_{sg}$ and $\beta_{sg}$). 
The procedure used for correcting $M_{1,A}$ is summarized in Algorithm \ref{alg:graddescent}. This algorithm also monitors
the normalized phase condition, $M_2=2\Re E_o^2[1+\cos(\theta_{d})]$. 
\begin{algorithm}
\caption{Iterative gradient descent method for minimizing risk function by optimizing biases and control signal gains.}
\begin{algorithmic}[1]
  \STATE Compute predicted $M_{1,P}$ with ideal $\alpha_{dc}$, $\beta_{dc}$, $\gamma$, $\alpha_{sg}$, $\beta_{sg}$
\STATE Read actual $M_{1,A}$
\STATE Compute the risk function $\mathbf{\textit{J}}_{[0]}(\alpha_{dc}$, $\beta_{dc}$, $\gamma$, $\alpha_{sg}$, $\beta_{sg})$

$\left(\mathbf{\textit{J}}_{[0]}=\dfrac{1}{N}\sum\limits_{i=1}^N (M_{1,A[0]}^{(i)}-M_{1,P}^{(i)})^2\right)$
\STATE Initialize $\alpha_{dc[0]}$, $\beta_{dc[0]}$, $\gamma_{[0]}$, $\alpha_{sg[0]}$, $\beta_{sg[0]}$
\IF {($\mathbf{\textit{J}}_{[0]}>0.001\%$)}
\FOR {$n = 0$ to $\text{epochs}$}
\STATE $\gamma_{[n+1]}=\gamma_{n}-\mu\times\nabla_{\gamma}\hspace{.1cm}\mathbf{\textit{J}}_{[n]}(\cdot)$
\STATE $\beta_{sg[n+1]}=\beta_{sg[n]}-\mu\times\nabla_{\beta_{sg}}\mathbf{\textit{J}}_{[n]}(\cdot)$
\STATE $\beta_{dc[n+1]}=\beta_{dc[n]}-\mu\times\nabla_{\beta_{dc}}\mathbf{\textit{J}}_{[n]}(\cdot)$
\STATE $\alpha_{sg[n+1]}=\alpha_{sg[n]}-\mu\times\nabla_{\alpha_{sg}}\mathbf{\textit{J}}_{[n]}(\cdot)$
\STATE $\alpha_{dc[n+1]}=\alpha_{dc[n]}-\mu\times\nabla_{\alpha_{dc}}\mathbf{\textit{J}}_{[n]}(\cdot)$
\STATE Compute $M_{1,A[n+1]}^{}$
\STATE 
$\mathbf{\textit{J}}_{[n+1]}=\dfrac{1}{N}\sum\limits_{i=1}^N (M_{1,A[n+1]}^{(i)}-M_{1,P}^{(i)})^2$
 \ENDFOR
    \ENDIF
\end{algorithmic}
\label{alg:graddescent}
\end{algorithm}

  \begin{figure}[tb!]
	\centering{
		\begin{tabular}{cccc}
\includegraphics[width=.22\textwidth]{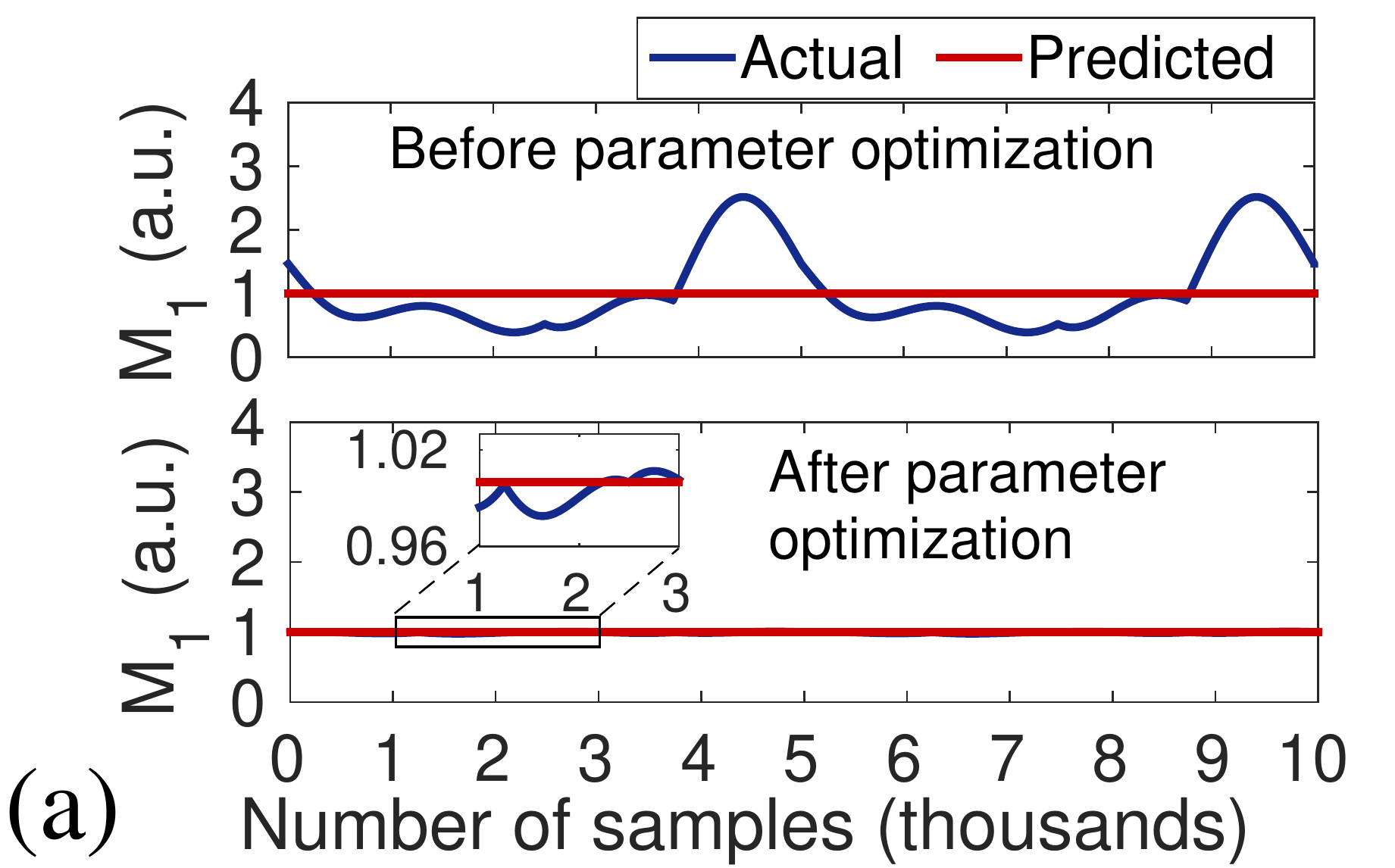}& \hspace{-.3cm}
\includegraphics[width=.22\textwidth]{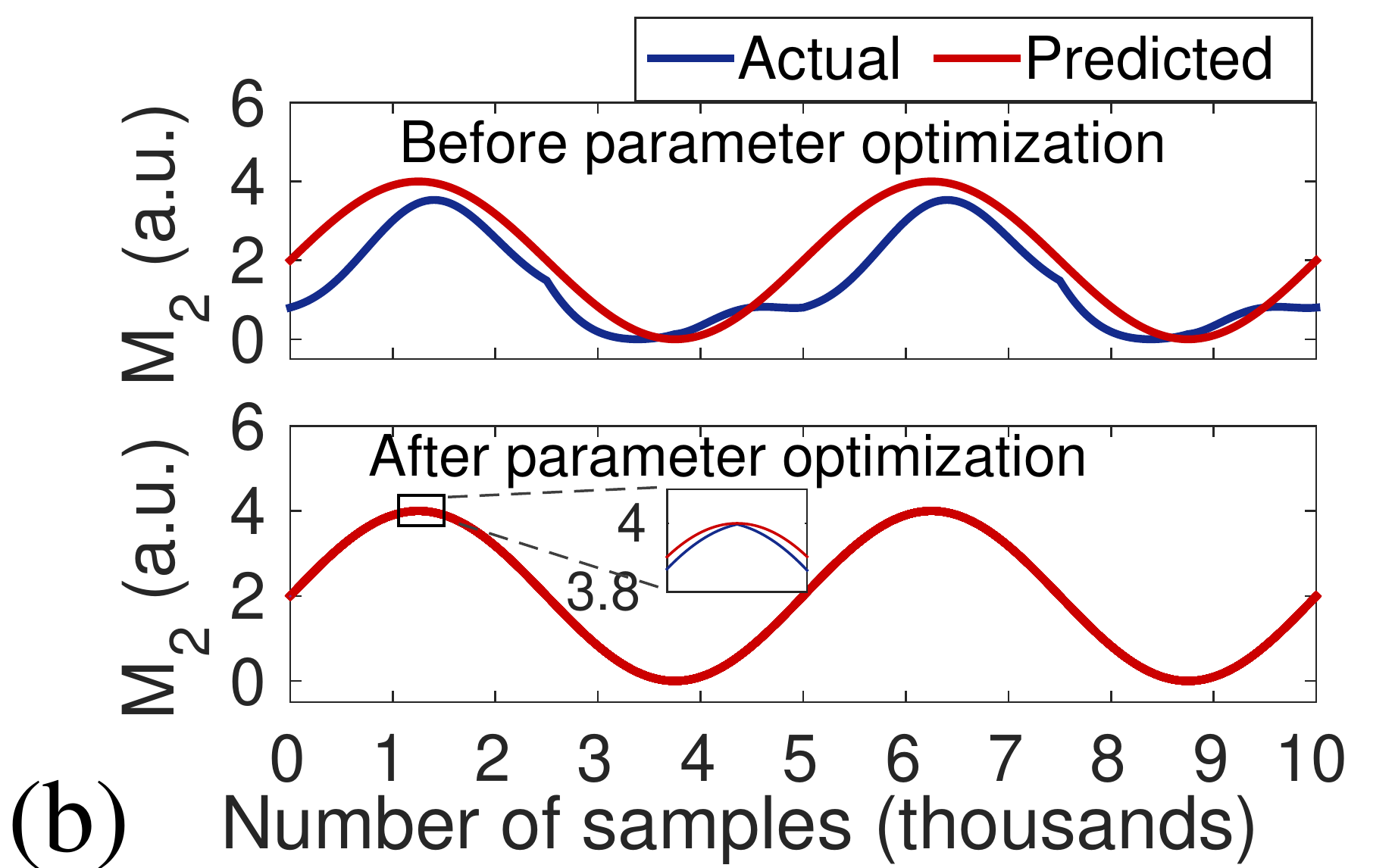}\\
\includegraphics[width=.22\textwidth]{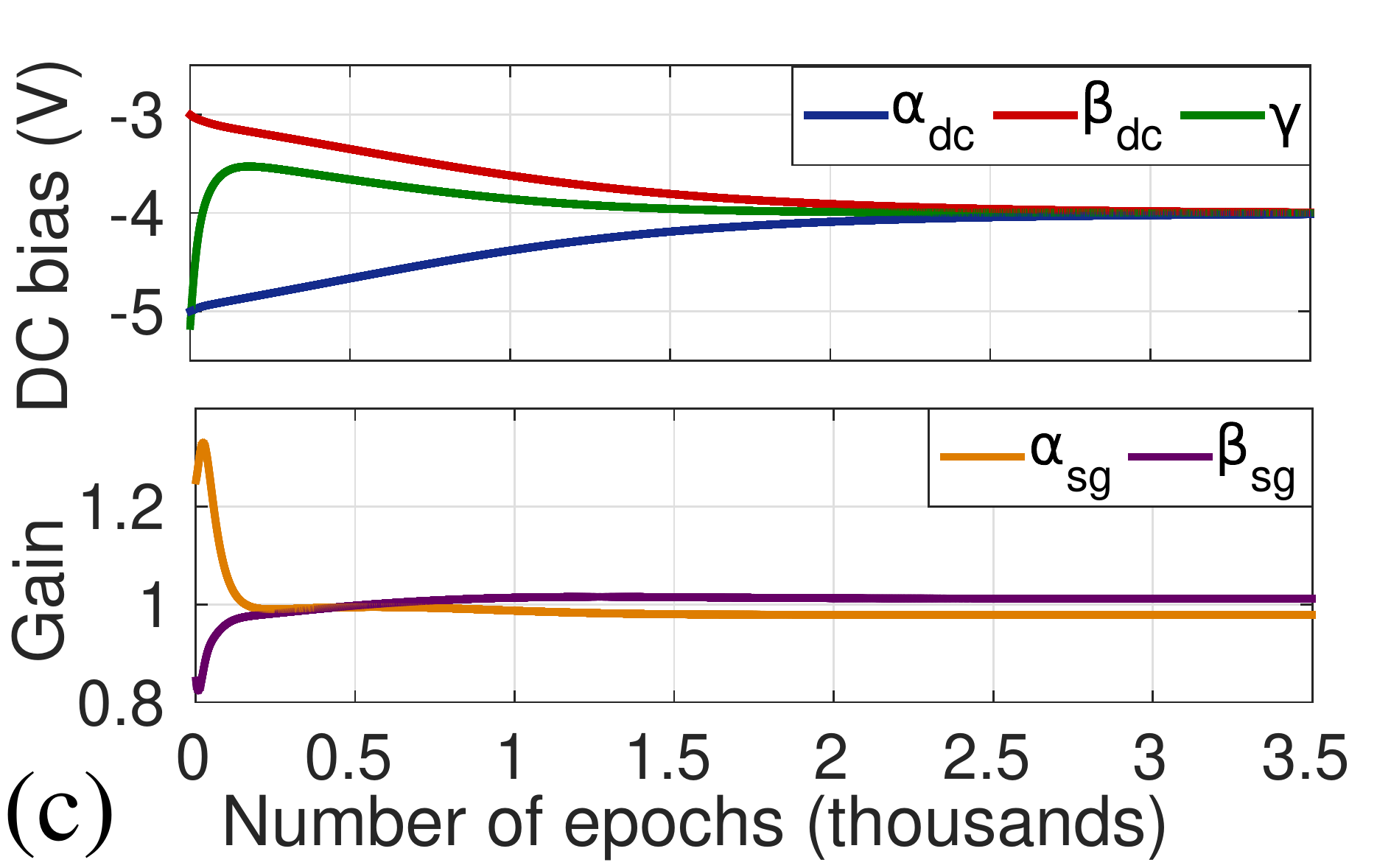}& \hspace{-.3cm}
\includegraphics[width=.22\textwidth]{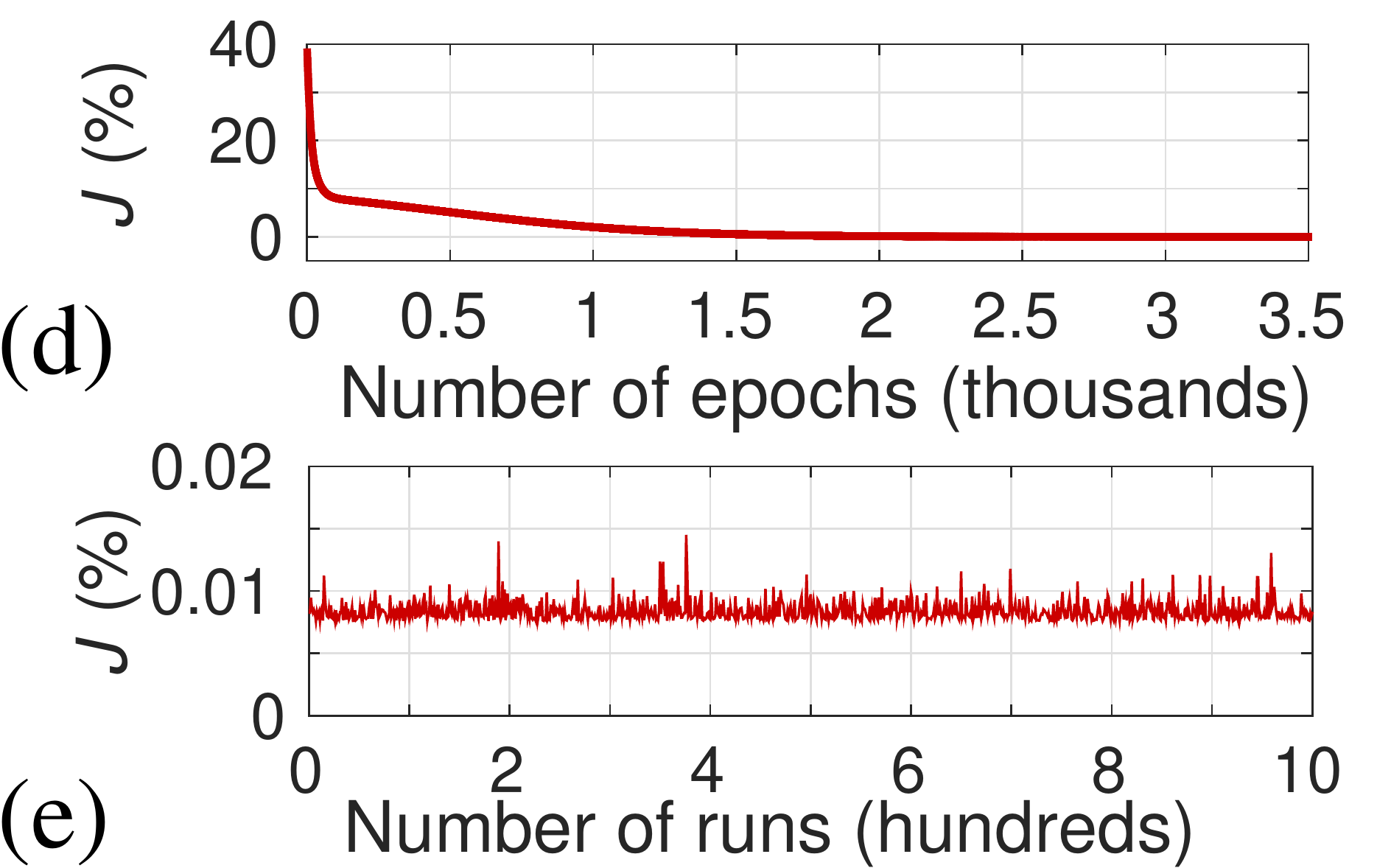}\\
\end{tabular}}
		\caption{Simulation results of bias voltages and control signals gain correction 
		using multi variate gradient descent algorithm. (a) Magnitude waveform ($M_1$) before 
		and after the parameter optimization; (b) Phase waveform ($M_2$) before 
		and after the parameter optimization; (c) Settling of bias voltages and control signal gains; 
		(d) Minimization of risk function for a single experiment; and (e) Settling of risk function for 
		one thousand experiments.}
		\label{fig:simresults}
	\end{figure}
Figure \ref{fig:simresults} presents the results obtained by optimizing the EOPD parameters. Due to fluctuations in the EOPD parameters, 
$M_{1,A}$ shows huge variations, deviating from $M_{1,P}$. After optimizing the parameters, 
$M_{1,A}$ approaches $M_{1,P}$ as shown in 
Fig. \ref{fig:simresults}(a). Similar behavior for $M_2$  is observed in Fig. \ref{fig:simresults}(b), {showing
proximate predicted and actual values. The zoomed version also emphasizes the deviation of actual value by $\sim$1\% 
from the predicted value.} {With the 
proposed architecture and algorithm, if the EOPD parameters are optimized, both $M_1$ and $M_2$ can be corrected, 
simulatanously.}
Corresponding settling of EOPD parameters is observed in Fig. \ref{fig:simresults}(c). Convergence 
of $\mathbf{\textit{J}}$ shown in Fig. \ref{fig:simresults}(d) signifies its minimization from 40\% 
to $<$0.1\%. The simulation is run for one thousand different cases of variations of $M_1$, and the minimization of 
$\mathbf{\textit{J}}$ is plotted in Fig. \ref{fig:simresults}(e), 
which validates this technique's feasibility for correcting $M_{1,A}$, in the presence of deviation of EOPD parameters within a range of $\pm$30\% 
from their nominal values. In certain cases, the risk function shows a higher order of magnitude, which is due to these parameters showing variations greater than $\pm$30\% 
from their nominal values. 
Such aberration is associated with $\mathbf{\textit{J}}$ as it is not strictly convex (verified using Hessian matrix, which
was not positive definite).
To resolve this issue, when $\mathbf{\textit{J}}$ crosses a certain threshold, 
the EOPD parameters are reset.

\begin{figure}[t!]
    \centering
    \includegraphics[width=.45\textwidth]{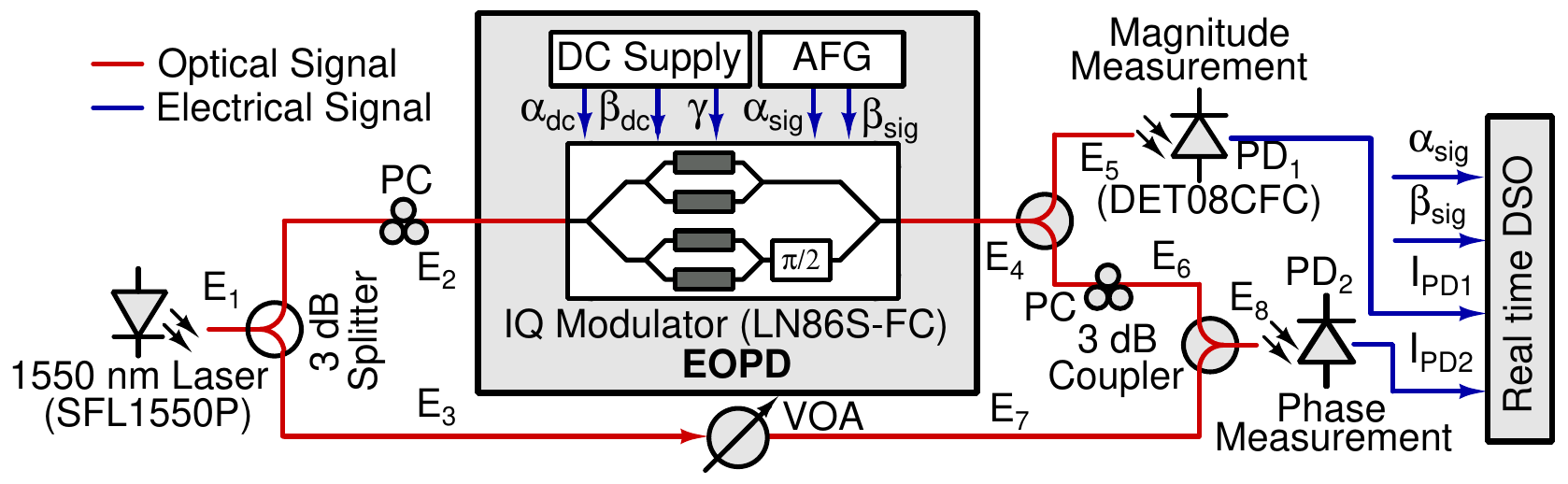} 
    \caption{Experimental set up for validating the endless optical phase delay. 
		PC: Polarization controller; AFG: Arbitrary waveform generator; VOA: Variable optical attenuator; PD: Photodetector;
		DSO: Digital storage oscilloscope.    }
    \label{fig:EOPD_TB}   
  \end{figure}

Experimental setup to validate the EOPD operation is shown in 
{Fig. \ref{fig:EOPD_TB}}. A laser output $E_1=\exp(j\omega_ot)$ is split to 
generate $E_{2}$ and $E_{3}$. $E_2$ is given to the IQ modulator, which receives 
biases and control signals from DC supply and 
arbitrary function generator (AFG), respectively. The $\alpha_{dc}$, $\beta_{dc}$, and $\gamma$ are set to $-V_{\pi}$, $-V_{\pi}$, and 
$V_{\pi}/2$, respectively with $\alpha_{sig}$ and $\beta_{sig}$ having peak to peak swing of $2V_{\pi}$ and a frequency of 
1\,MHz.  In this configuration, 
the combination of IQ modulator and CE acts as EOPD generating $E_4=\exp[j(\omega_ot+\theta_d)]$. $E_4$ is equally split
separately for validating the magnitude and the phase conditions to give out $E_5$ 
and $E_6$. $E_5$ fed to PD$_1$ generates 
photocurrent $I_{PD1}=\Re|E_5|^2$ and is converted into $V_{PD1}=I_{PD1}\times50\Omega$. $E_6$ is combined with $E_4$ 
and then given to PD$_2$ forming an interferometer. The interferometer output $I_{PD2}=2[1+\cos(\theta_d)]$ is converted into 
voltage using 50\,$\Omega$ load. These waveforms are observed on an oscilloscope.

\begin{figure}[tb!]
	\centering{
		\begin{tabular}{cccc}
\hspace{-.1cm}\includegraphics[width=.25\textwidth]{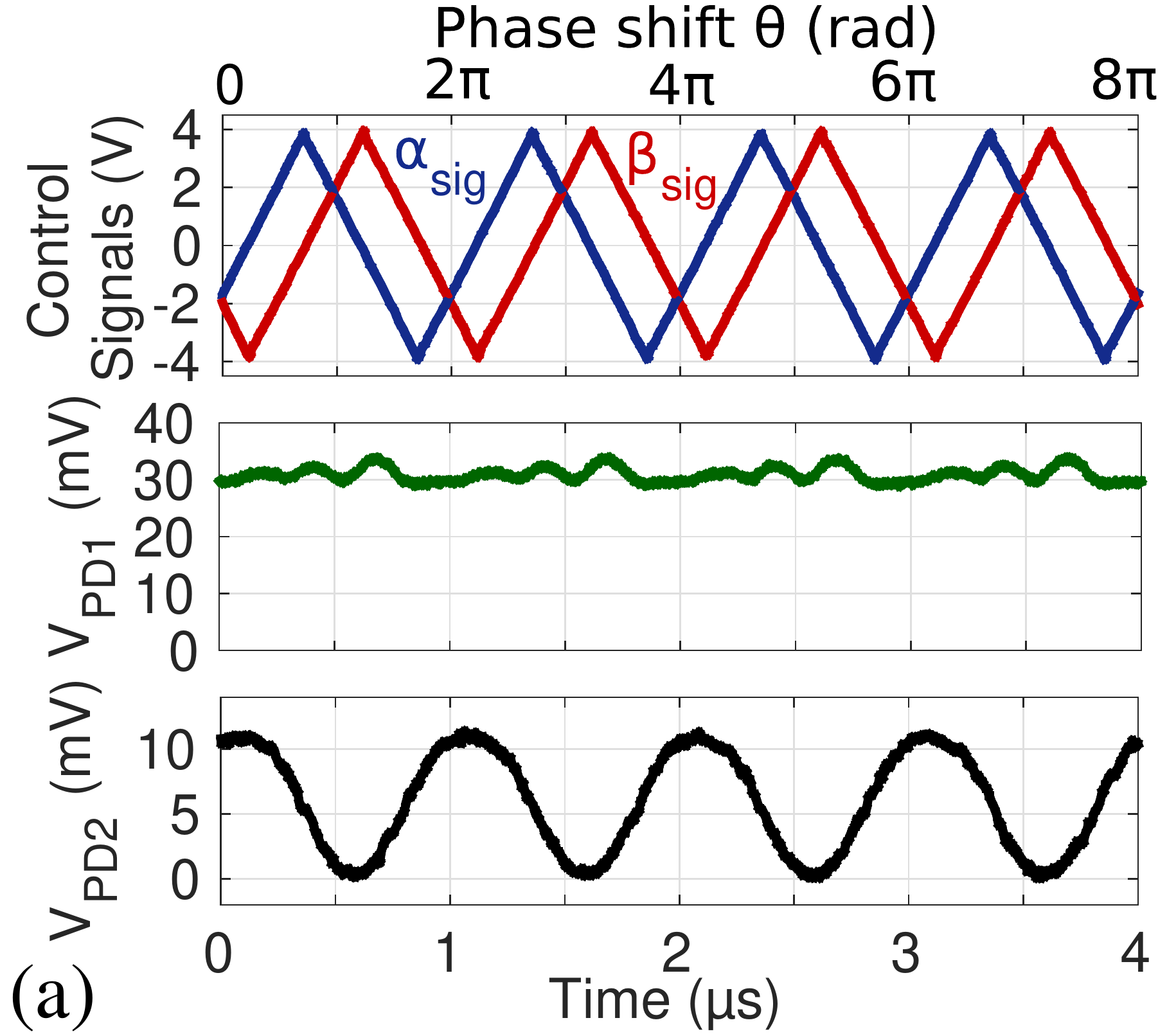}& 
\hspace{-.2cm}\includegraphics[width=.195\textwidth]{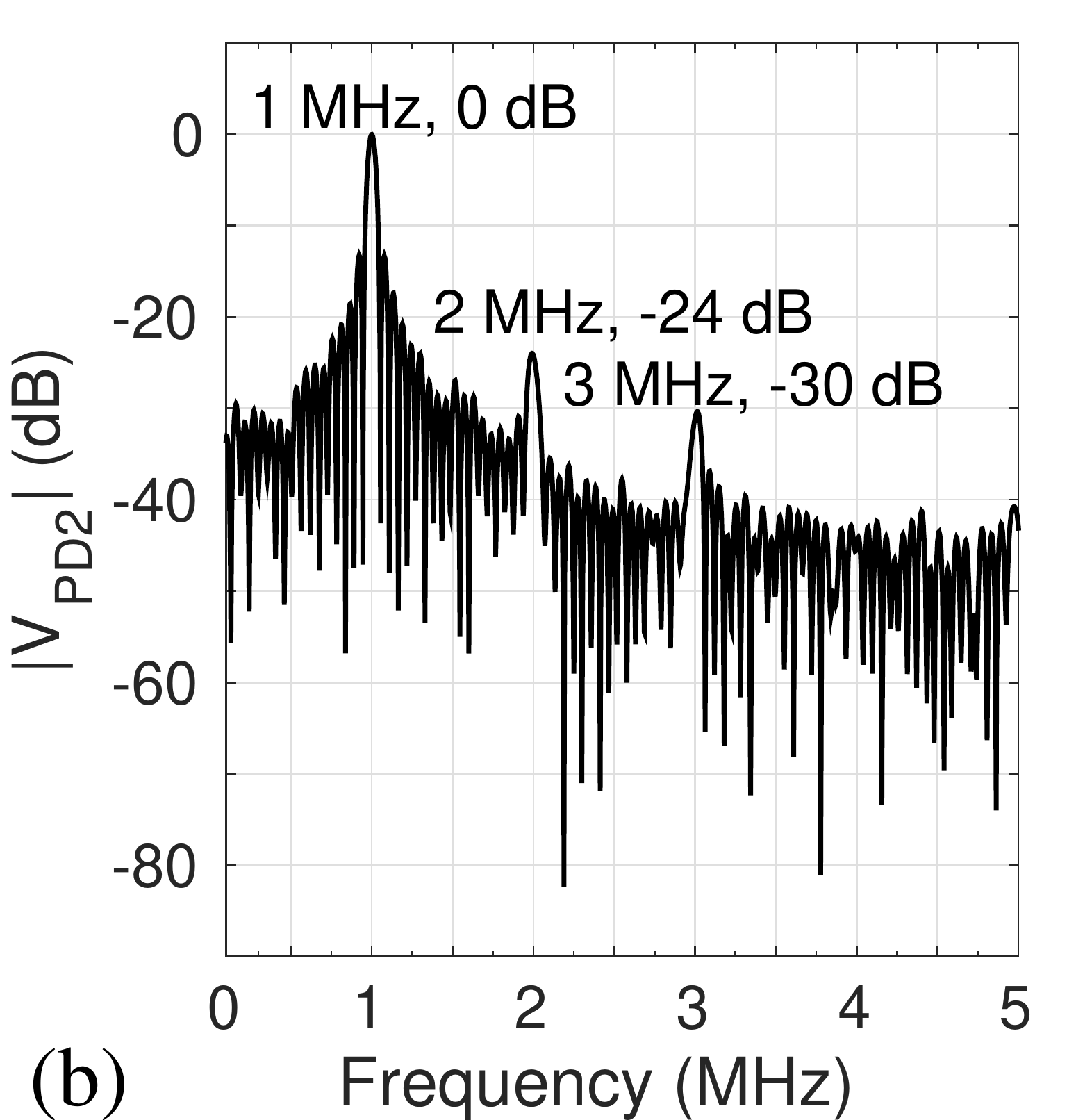}\\
\end{tabular}}
		\caption{ Experimental characterization results of endless optical phase delay. 
		(a) Time domain waveforms. $\alpha_{sig}$ and $\beta_{sig}$ are control signals, 
		$V_{PD1}$ and $V_{PD2}$ are signals validating magnitude and phase conditions; and
		(b) {Frequency spectrum of $V_{PD2}$}.}		
		\label{fig:EOPD_results}
	\end{figure}

	Experimental results of EOPD are shown in {Fig. \ref{fig:EOPD_results}}. Control signals of 1\,MHz 
	 {
	introduce a phase delay to the input signal at a rate of $2\pi$\,rad$/$\si{\micro}s.}	
	The $V_{PD1}$ waveform, signifying the behavior of magnitude condition, is approximately constant. The interferometer output 
	$V_{PD2}$ is a 1\,MHz sinusoidal signal {signifying
	the addition of a phase delay at a rate of $2\pi$\,rad/\si{\micro}s over time. This behavior is similar to phase accumulation
	at a rate of $2\pi$ per cycle as observed in Fig. \ref{fig:EOPD_sim_results}}. 
	The frequency spectrum of $V_{PD2}$ depicted in {Fig. \ref{fig:EOPD_results}(b)}, which shows that the fundamental frequency is 
	{at least} 20\,dB above 
	its harmonics validating phase condition. The EOPD's operation conditions can be improved further by incorporating a lookup 
	table in the algorithm to take care of second order effects in the modulator.
	
\section{EOPD for Phase Synchronization in a DCI}
\noindent The EOPD can be used for many applications, such as optical frequency shifting \cite{Izutsu_JQE1981}, 
MIMO demultiplexing \cite{Doerr_USPatent2014}, multi-carrier generation \cite{Yamazaki_JSTQE2013}, RF sinusoidal signal 
generation using optical phase locked loops \cite{Balakier_JSTQE2018}, and carrier phase synchronization in 
coherent homodyne and self-homodyne links \cite{Ashok_CLEO2020}.

\subsubsection*{EOPD aided phase synchronization}
In PMC-SH links, the modulated signal ($S$) and the unmodulated carrier ($LO$) co-propagate along the same channel in two orthogonal polarizations and are separated at the receiver for demodulation \cite{Kamran_JLT2020}. Due to the path length difference, laser linewidth, and polarization mode dispersion, time varying phase offset is introduced between $S$ and $LO$, 
which has to be removed using phase synchronization. Analog signal processing based phase synchronization technique with the aid of conventional PM is
demonstrated in \cite{Ashok_OFC2019}. However, due to the bounded phase delay provided by conventional PM, the loop fails to track the phase offset when it exceeds a few radians. 
Under such conditions, EOPD is a solution to synchronize the phases of $S$ and $LO$.

\begin{figure}[t!]
    \centering
    \includegraphics[width=.48\textwidth]{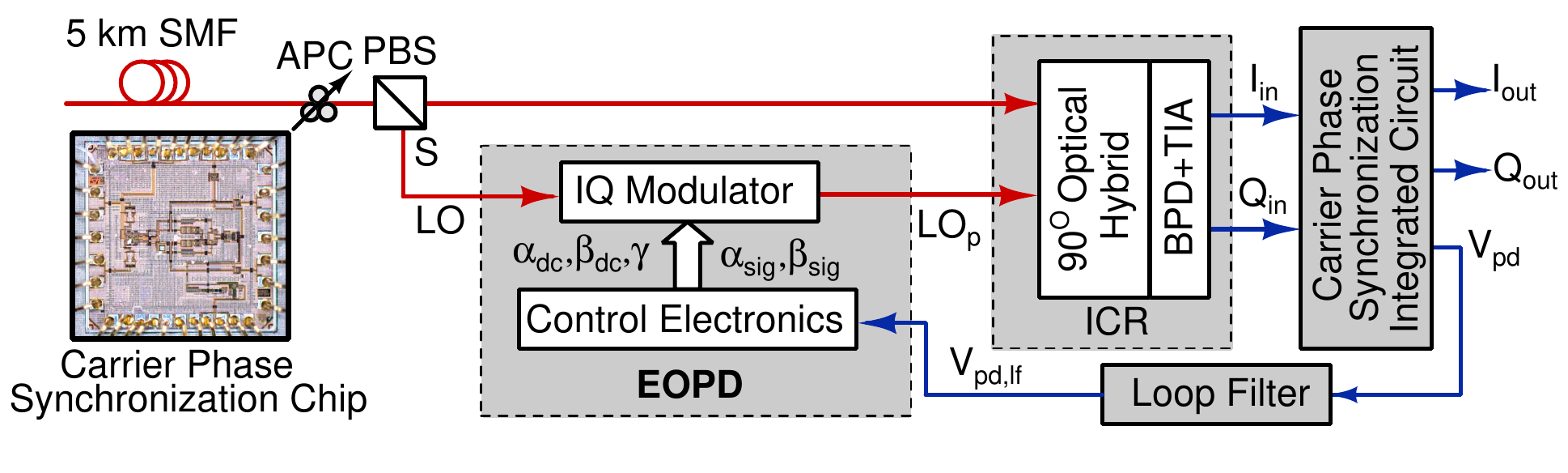} \vspace{-.1cm}
    \caption{Experimental setup of endless optical phase delay aided  phase synchronization in polarization multiplexed 
    carrier based self-homodyne QPSK receiver. SMF: Single mode fiber; APC: Adaptive polarization control; PBS: Polarization beam splitter; 
    $S$: Received modulated signal; $LO$, $LO_p$: Local oscillator 
		signal before and after phase shifting; ICR: Integrated coherent receiver; BPD: Balanced photodetectors; TIA: Trans-impedance amplifier; 
		$I_{in}$, $Q_{in}$: In-phase and quadrature-phase input signals; 
		$I_{out}$, $Q_{out}$: In-phase and quadrature-phase output signals; and $V_{pd}$, $V_{pd,lf}$: Phase detector output 
		before and after loop filter.}
		\vspace{-.1cm}
    \label{fig:EOPD_SH_CPRC_exp_setup}   
  \end{figure}

An experimental setup of the PMC-SH system with EOPD based phase synchronization is shown in Fig. \ref{fig:EOPD_SH_CPRC_exp_setup}.
At the receiver side of the PMC-SH link, $S$ and $LO$ are obtained after adaptive polarization control (APC). The EOPD takes $LO$ 
and gives out a phase delayed version $LO_p$. 
The $S$ and $LO_p$ are 
mixed and down-converted to baseband in-phase and quadrature-phase electrical signals 
$I_{in}=\cos[\phi_{m}(t)+\phi_{off}(t)]$ and $Q_{in}=\sin[\phi_{m}(t)+\phi_{off}(t)]$, where $\phi_m$ and $\phi_{off}$ denote 
message phase and 
time varying phase offset, respectively. These signals are given to analog domain proof-of-concept phase synchronization chip 
 that generates phase detector output $V_{pd}$, which is proportional 
to $\phi_{off}$. The $V_{pd}$ is given to the EOPD through loop filter. The CE generates bias voltages and 
control signals corresponding to the $V_{pd,lf}$. EOPD adds phase delay to $LO$ to generate $LO_p$ based on the control signals' 
phase/frequency. At steady state, in closed loop condition, the phases of control signals are such that the phase delay added to 
$LO$ is equal and opposite to $\phi_{off}$ resulting in phase alignment of $S$ and $LO_p$ and hence obtaining 
$I_{out}=\cos[\phi_{m}(t)]$ and $Q_{out}=\sin[\phi_{m}(t)]$, which are phase offset corrected signals obtained at the 
output of the phase synchronization chip.

\begin{figure}[tb!]
	\centering{
		\begin{tabular}{cccc}
\includegraphics[width=.19\textwidth]{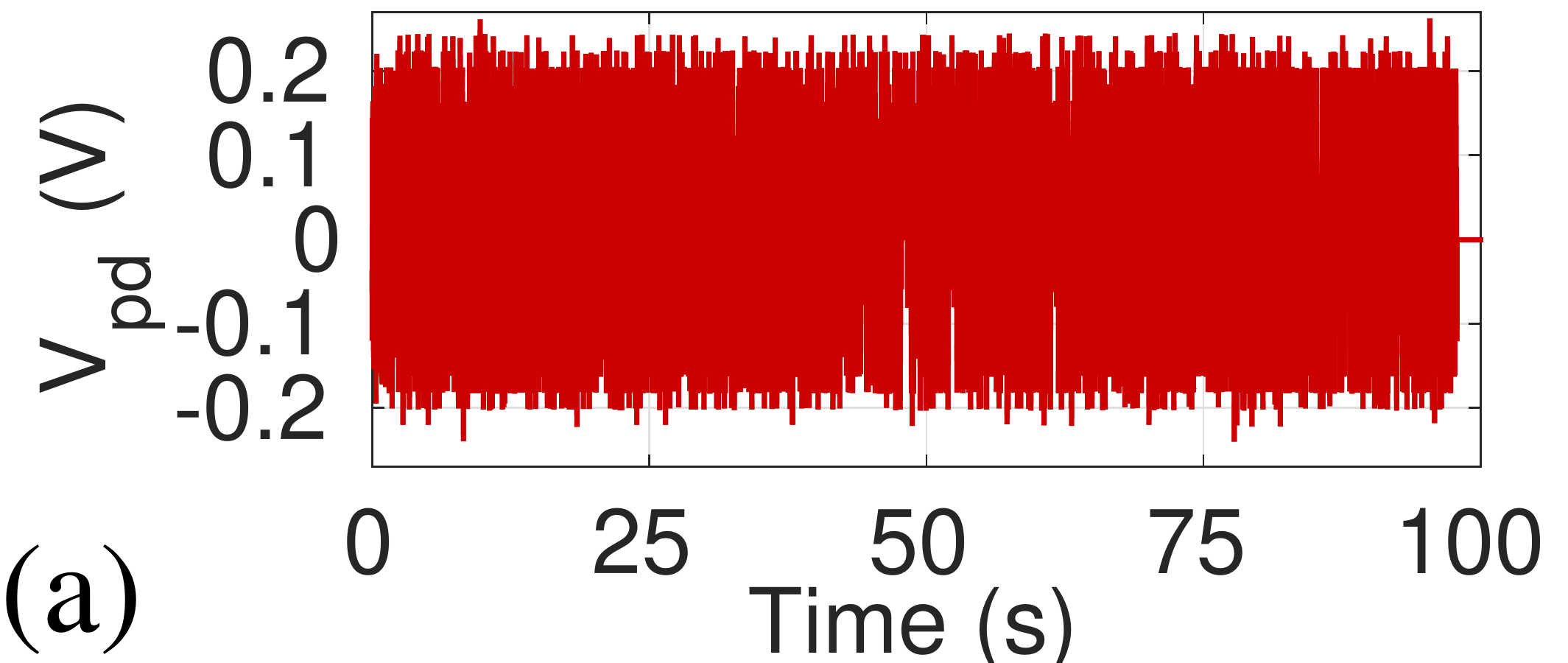}& 
\includegraphics[width=.19\textwidth]{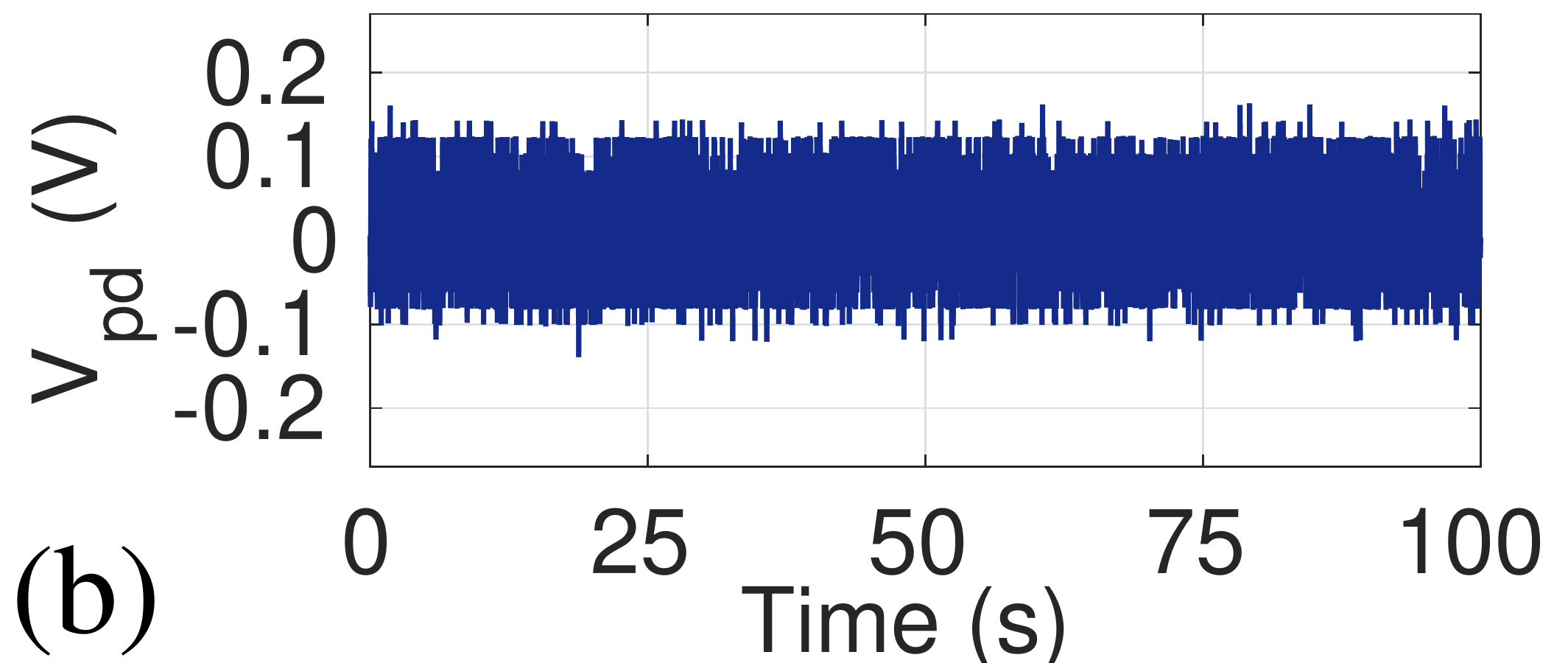}\\
\includegraphics[width=.21\textwidth]{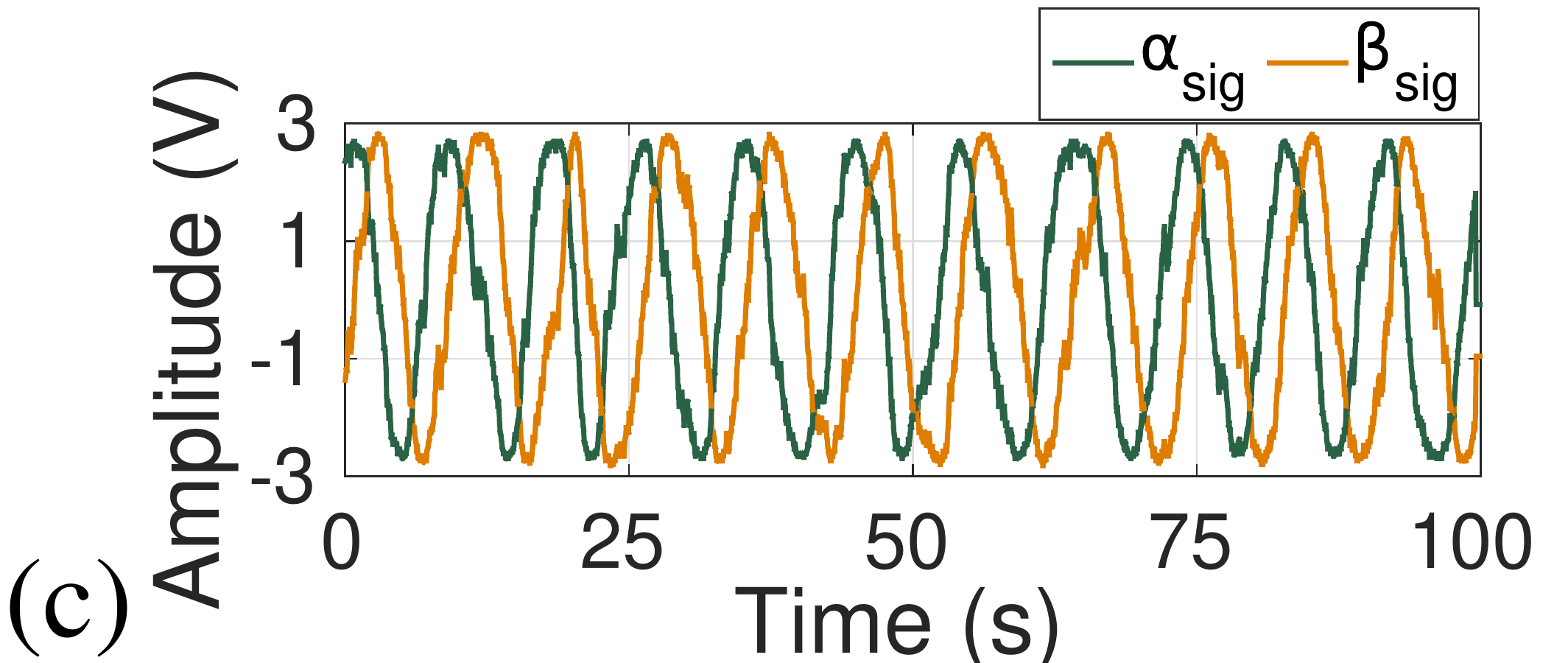}& 
\includegraphics[width=.21\textwidth]{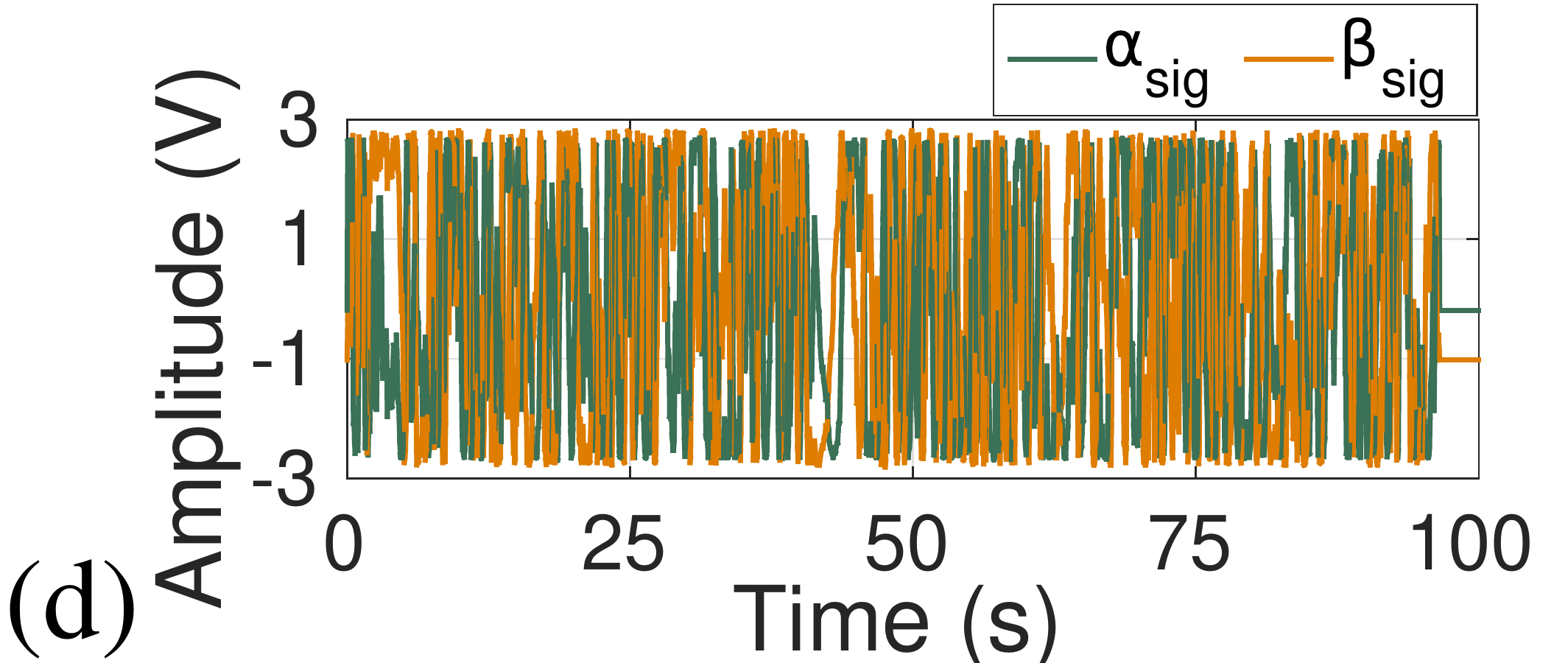}\\
\includegraphics[width=.22\textwidth]{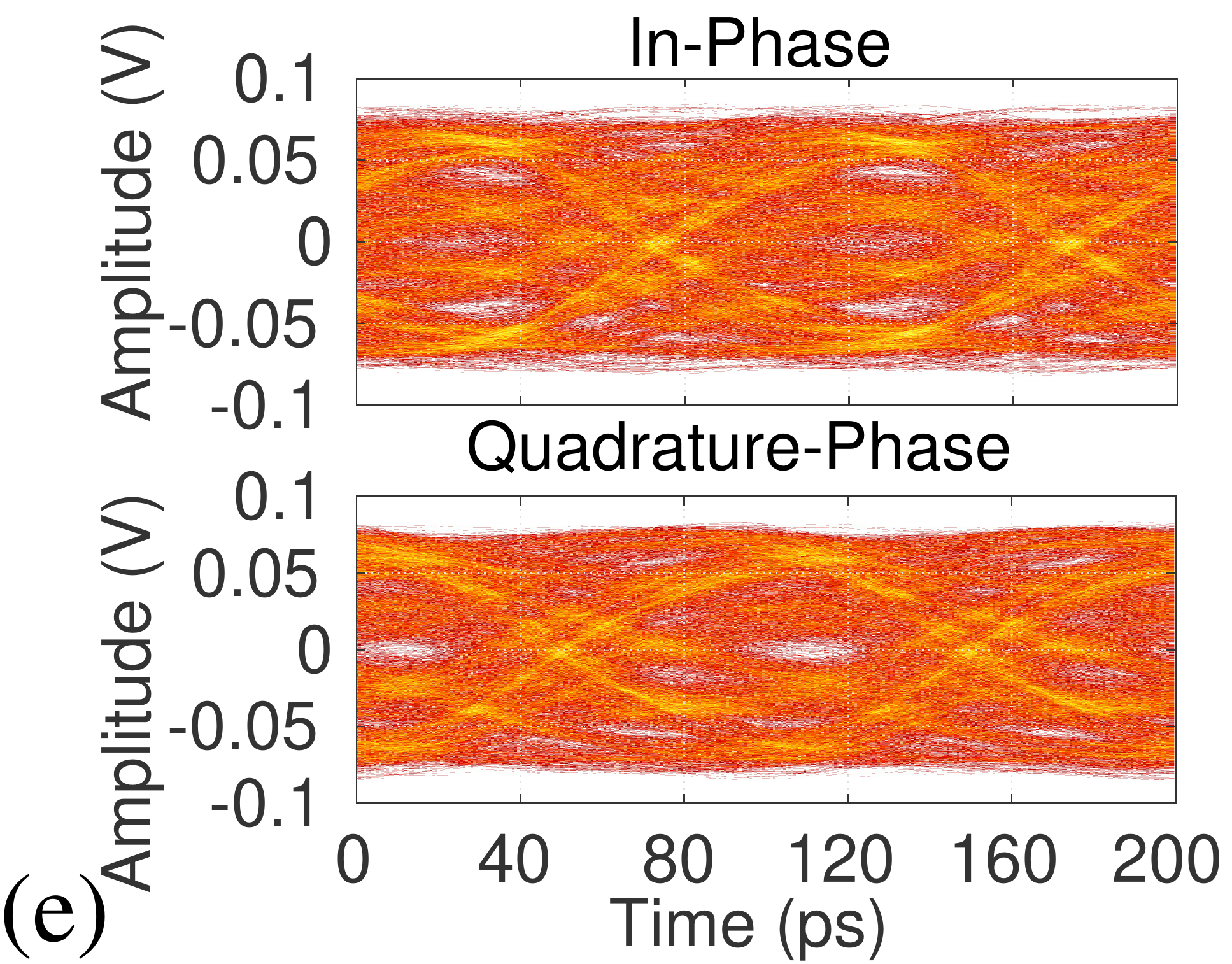}& 
\includegraphics[width=.22\textwidth]{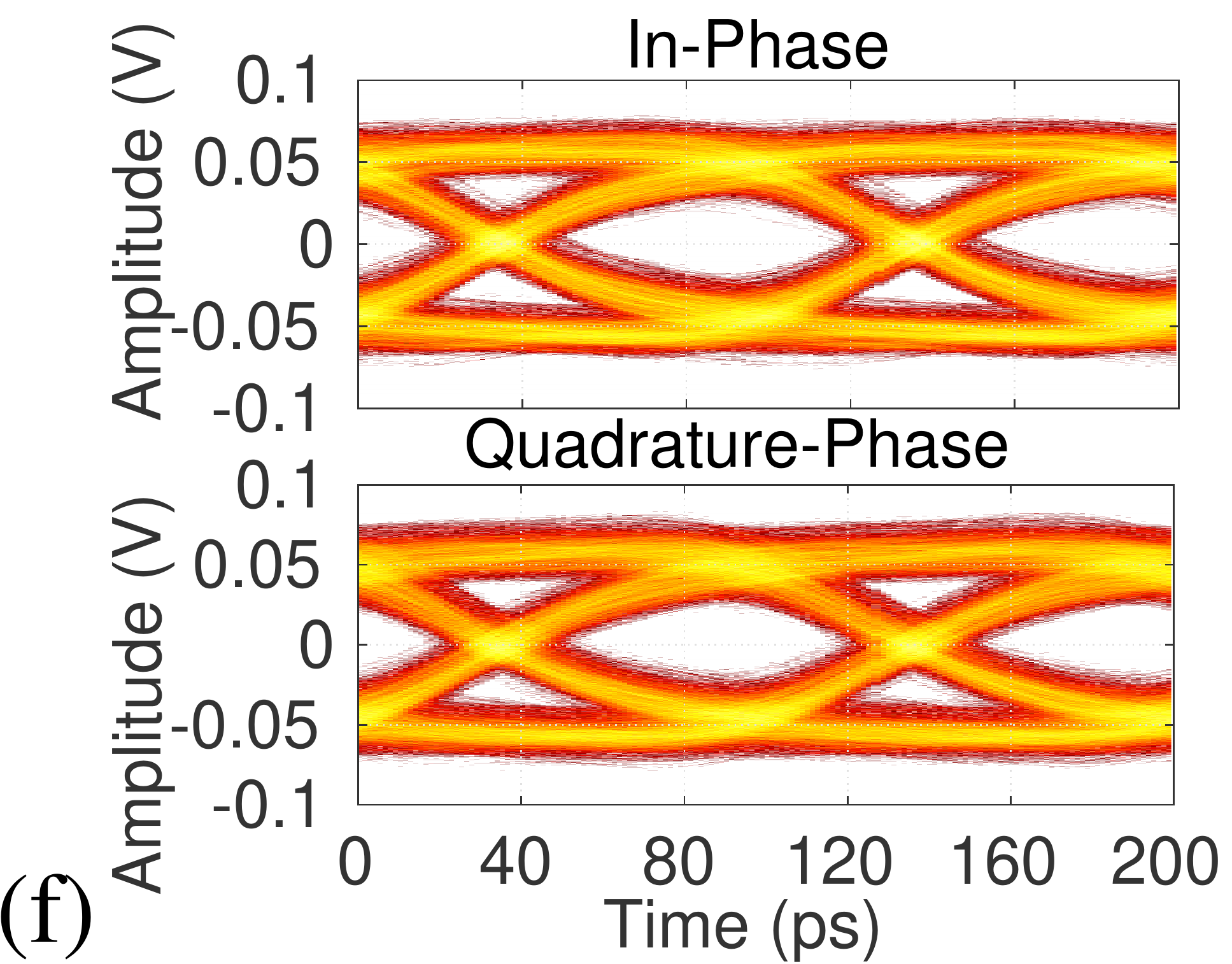}\\
\end{tabular}}
		\caption{ Experimental results obtained for 10\,GBaud, self-homodyne, 5\,km  QPSK link. 
		Phase detector outputs: (a) Open loop; and (b) Closed loop; Control signals: 
		(c) Open loop; and (d) Closed loop;
		Eye-diagrams of IQ signals: (e) Open loop; and (f) Closed loop.}		
		\label{fig:SH_CPRC_EOPD_results}
	\end{figure}

  Experimental results obtained with 20\,Gbps PMC-SH QPSK system are presented in Fig. \ref{fig:SH_CPRC_EOPD_results}. In open loop 
  condition, $V_{pd}$ shows a swing of $\sim$400\,mV$_\text{pp}$ (shown in
  Fig. \ref{fig:SH_CPRC_EOPD_results}(a)) and 
  corresponding EOPD control signals are triangular, as shown in Fig. \ref{fig:SH_CPRC_EOPD_results}(c). As a result, the in-phase and 
  quadrature-phase signals show closed eye-diagrams, as shown in Fig. \ref{fig:SH_CPRC_EOPD_results}(e). In closed loop 
  condition, the $V_{pd}$ signal shows lower amplitude (shown in Fig. \ref{fig:SH_CPRC_EOPD_results}(b)) resulting in 
  continuously time varying control signals (shown in Fig. \ref{fig:SH_CPRC_EOPD_results}(d)). Such behavior of $V_{pd}$, $\alpha_{sig}$, and 
  $\beta_{sig}$ signify continuous tracking of phase offset by the loop. Closing the loop results in opening of the eye, as shown in Fig. 
  \ref{fig:SH_CPRC_EOPD_results}(f) due to phase offset correction. {The proposed EOPD aided carrier phase synchronization 
  is independent of data rate as EOPD aims at correcting $\phi_{off}(t)$. This proof-of-concept demonstration
  is done at 20\,Gbps due to frequency limitations posed by the analog domain carrier phase synchronization chip and the PCB assembly.
  The power dissipation of the proposed EOPD can range from less than a milliwatt to a few tens of milliwatts depending on the speed of phase shift that has to be changed, the type of phase shifters, and electronics used.}
  
\section{Conclusion}
\noindent We presented an EOPD that can add an arbitrary amount of phase delay to an optical signal.  The EOPD can be tuned seamlessly and achieve an infinite delay range. 
The EOPD can be used in many practical applications, including phase/frequency shifting in coherent links and DCIs 
(one such application was presented). The EOPD can also be employed in LiDAR and multiple phased antenna array receiver
applications.

\ifCLASSOPTIONcaptionsoff
  \newpage
\fi

\bibliographystyle{IEEEtran}
\bibliography{references_rakesh_v1.bib}

\end{document}